\documentclass[journal,twoside,web]{ieeecolor}
\usepackage{tmi}
\usepackage{amsmath,amssymb,amsfonts}
\usepackage{algorithmic}
\usepackage{booktabs}
\usepackage{cite}
\usepackage{color,xcolor}

\usepackage{graphicx}
\usepackage[backref]{hyperref}
\usepackage[utf8]{inputenc}
\usepackage{longtable}
\usepackage{makecell}
\usepackage{mathrsfs}
\usepackage{multicol}
\usepackage{multirow}
\usepackage{textcomp}
\usepackage{url} 

\def\BibTeX{{\rm B\kern-.05em{\sc i\kern-.025em b}\kern-.08em
    T\kern-.1667em\lower.7ex\hbox{E}\kern-.125emX}}
\begin{document}
\title{Realistic Restorer: artifact-free flow restorer\\ (AF2R) for MRI motion artifact removal}
\author{Jiandong~Su,~Kun~Shang,~and~Dong~Liang,~\IEEEmembership{Senior Member, IEEE}
\thanks{This work was supported in part by the National Natural Science Foundation of China under Grant 12001180, and Grant 12026603. (Jian-dong~Su~and~Kun~Shang are contributed equally to this work, Corresponding author is Dong~Liang.)}
\thanks{Jian-dong~Su,~Kun~Shang~and~Dong~Liang are with the Research Center for Medical AI, Shenzhen Institutes of Advanced Technology, Chinese Academy of Sciences, Shenzhen, Guangdong CN. Dong~Liang is also affiliated with Pazhou Lab, Guangzhou, CN. (e-mail: jd.su@siat.ac.cn; kun.shang@siat.ac.cn; dong.liang@siat.ac.cn)}}

\maketitle

\begin{abstract}
 Motion artifact is a major challenge in magnetic resonance imaging (MRI) that severely degrades image quality, reduces examination efficiency, and makes accurate diagnosis difficult. However, previous methods often relied on implicit models for artifact correction, resulting in biases in modeling the artifact formation mechanism and characterizing the relationship between artifact information and anatomical details. These limitations have hindered the ability to obtain high-quality MR images. In this work, we incorporate the artifact generation mechanism to reestablish the relationship between artifacts and anatomical content in the image domain, highlighting the superiority of explicit models over implicit models in medical problems. Based on this, we propose a novel end-to-end image domain model called AF2R, which addresses this problem using conditional normalization flow. Specifically, we first design a feature encoder to extract anatomical features from images with motion artifacts. Then, through a series of reversible transformations using the feature-to-image flow module, we progressively obtain MR images unaffected by motion artifacts. Experimental results on simulated and real datasets demonstrate that our method achieves better performance in both quantitative and qualitative results, preserving better anatomical details.
\end{abstract}


\section{Introduction}\label{sec:introduction}
\IEEEPARstart{M}{agnetic} Imaging (MRI) is one of the most commonly used medical imaging techniques in clinical practice. MRI allows flexible and non-invasive mapping of anatomical structures without the need for ionizing radiation. Additionally, MRI offers the advantage of being able to select desired planes by adjusting the magnetic field, providing images of areas that may be inaccessible or difficult to approach using other imaging modalities. However, MRI also has its limitations. One of the major drawbacks is the long scanning time~\cite{wood1985mr,stadler2007artifacts}. This prolonged duration increases the possibility of patient motion, leading to inaccurate acquisition of physiological signals and resulting in artifacts in the obtained images. These artifacts can make it hard to observe tissue structures and decrease the diagnostic value of the images, ultimately leading to ineffective examinations and low diagnostic efficiency.

Artifacts are abnormal signals caused by external factors that do not exist in the original physiological signal. The manifestation of artifacts in the image domain depends on the type and intensity of the external signals. Common external signals that cause artifacts is patient's motion. Voluntary or involuntary motion by the patient leads to a significantly shorter sampling time for frequency-encoding direction than a phase-encoding time, resulting in phase disturbances.
Patient's motion can be categorized into rigid motion and elastic motion. Previous studies have mostly focused on rigid motion and have not adequately distinguished between the patient's rigid motion and elastic motion. Therefore, our research primarily focuses on respiratory motion, which is a form of elastic motion. 



Many studies have previously addressed this issue from the signal acquisition stage. Some researcher\cite{ehman1984magnetic,pipe1999motion,welch2002spherical,white2010promo,qin2009prospective} proposed the use of hardware modifications to estimate motion and subsequently apply motion correction based on the obtained motion estimates.Such as PROPELLER navigator~\cite{pipe1999motion}, spherical navigators \cite{welch2002spherical}, and so on. Those k-space based navigators require accurate and consistent k-space sampling trajectories between acquisitions, which would  be violated by several factors, e.g., gradient instabilities and magnetic field inhomogeneities~\cite{white2010promo}. Additionally, some research attempts to tackle this problem from a sequence perspective\cite{vasanawala2010navigated,chavhan2013abdominal}. However, the mentioned methods require additional hardware, which incurs high costs.

In order to reduce costs, some researchers have attempted to address this issue during the reconstruction period. Reconstruction-based methods primarily aim to incorporate motion trajectory prediction\cite{haskell2018targeted,haskell2019network}, motion parameter estimation\cite{polak2022scout}, motion estimation\cite{loktyushin2013blind}, compressed sensing\cite{haldar2013low,haldar2016p,kim2017loraks}. However, the presence of artifacts in MR images is determined by evaluating the reconstructed images. This means that removing artifacts from a reconstruction perspective requires multiple reconstructions, leading to a significant decrease in efficiency. Additionally, hospitals rarely store raw data due to storage limitations, while reconstruction-based methods rely on raw data, further complicating their practical application.

With the remarkable emergence of deep learning, image-domain methods based on deep learning have gained considerable attention~\cite{pawar2018motion,duffy2021retrospective,shaw2020k,scannell2019robust}. One common approach is using deep CNNs to estimate the motion between the motion artifact image and the reference image, and then correct it~\cite{zhang2019multi,liu2020motion,tamada2020motion}, which treat artifact removal as a image restoration task. The other is converting the task to image to image translation (I2IT) with deep generative model. Armanious~\textit{et al.} proposed MedGAN~\cite{armanious2020medgan}  based on deep generative adversarial network~(GAN)~\cite{goodfellow2014generative} for medical image translation, which shows good performance for motion artifact correction. And he~\cite{armanious2019unsupervised} further suggested a unsupervised translation framework, named cycle-MedGAN, to utilize the non-adversarial cycle losses which direct the framework to minimize the textural and perceptual discrepancies in the translated images without pair data. Oh~\textit{et al.}~\cite{oh2021unpaired} proposed another unpaired GAN-based approaches with bootstrap aggregation that does not require paired motion-free and motion artifact images from the reconstruction perspective. Upadhyay~\textit{et al.}~\cite{upadhyay2021uncertainty} proposed an uncertainty-guided progressive learning scheme by incorporating aleatoric uncertainty as attention maps for GANs trained in a progressive manner for a holistic view. 

As illustrated above, it is a more widely accepted view to solve this problem using GAN. Due to idea of adversarial learning and good expression ability, GAN is under heated discussion by researchers. However, due to GAN's implicit modeling, the generated samples of GAN is lack of control. Due to the nature of the modeling approach, implicit generative models always generate approximate values rather than deterministic ones. This limitation arises when using such models to solve problems that require high precision. Therefore, implicit generative models is quite different from the fundamental requirement of medical image fidelity. Therefore, designing a generation model with reliable generation results and excellent expression ability is worthy of study.
 
In this paper, 
we introduce AF2R, a new deep model based on normalizing flow for motion artifact removal. Compared with the GANs, the proposed method can directly model the distribution of data and assist in generation by explicit modeling the training data. Besides, our models have an explicit generative process, which allows for better control over generated samples. In tasks that require high fidelity and fine-grained details, such as medical image quality control, our models can better capture subtle variations in the data distribution and generate more realistic and accurate samples. In addition, by incorporating the mechanism of artifact generation, we reformulate the relationship between artifact and anatomical structure information in the image domain. Based on this, we redesign the coupling layers of normalizing flows. Our subsequent experiments confirmed this. 

To sum up, we highlight the main contribution of this work as follows: 
\begin{itemize}
  \item{We demonstrate the necessity of explicit modeling approach of image domain-based method in medical imaging, which has been the main focus of previous studies.}
  \item{We redefine the relationship between artifact and anatomical structure in the image domain and propose a new nonlinear relationship between them.}
  \item{To best of our knowledge, we introduce normalizing flow into the quality control task of MR images for the first time, and propose a novel method for motion artifact removal based on normalizing flow. In both simulated and real scenarios, our method shows better performance than contemporaneous work.}
  
\end{itemize}
\section{BACKGROUND AND MOTIVATION}
Motion artifacts in MR images refer to unwanted image distortion or blurring caused by movement of the patient during the image acquisition process.  When there exist subject’s motions during the frequency encoding, the displacement incurs K-space phase error at the specific phase encoding line. Obviously, it is caused by the corrupted part of K-space. 

\subsection{Reformulate Question}

From the perspective of respiratory motion correction(RMC), many previous studies\cite{jin2017mri,zhang2022flexible} try to solve this problem from the perspective of image reconstruction. The reconstruction of MR image can be essentially solving an inverse problem, which can be expressed as follows:
\begin{equation}
    f = Am+\varepsilon
\end{equation}
where $f$ is the collected K-space data,$A$ is the coding matrix, $m$ is the image to be reconstructed,and $\varepsilon$ represents the system noise. If respiratory movement incur, the signal acquired after respiratory movement $s'(k_x,k_y)$ can be obtained as follows:
\begin{equation}
s'(k_x,k_y)\!=\! \int\!\!\!\!\int\! \rho(x-p(k_y),y-q(q_y))e^{-2\pi({k_x}x+{k_y}y)}dxdy
\end{equation}
set $x'= x-p(k_y)$, $y = y-q(k_y)$,and get:
\begin{equation}
    s'(k_x,k_y)=\int\int\rho(x',y')e^{-2\pi({k_x}x+{k_y}y)}dxdy
\end{equation}
namely:
\begin{equation}
    s'(k_x,k_y)=e^{-i*\phi({k_x}+{k_y})}*s(k_x,k_y)\\
\end{equation}
where $\phi({k_x},{k_y})$ = $2\pi({k_x}p(k_y)+{k_y}q(k_y))$, $\rho(x',y')$ is the image obtained when the $k_y th$ K-space line is collected, which means that the effect of patient's movement is producing a phase disturbance in K-space. 

As for different motion catagories, rigid motion refers to the movement of an object where its shape and size remain unchanged. In rigid motion, the relative positions and distances between different parts of the object remain constant, and no deformation occurs. Examples of rigid motion include the movement of a patient's head or limbs. In contrast, elastic motion refers to the deformation of an object under external forces, which returns to its original shape after the force is removed. In elastic motion, the object undergoes deformation but can recover within a certain range. Common examples of elastic motion in the context of MRI include cardiac pulsation and respiratory motion. Generally, rigid motion is simpler compared to elastic motion and can be handled well. 

\begin{figure}
    \centering
    \includegraphics[width=11.5cm]{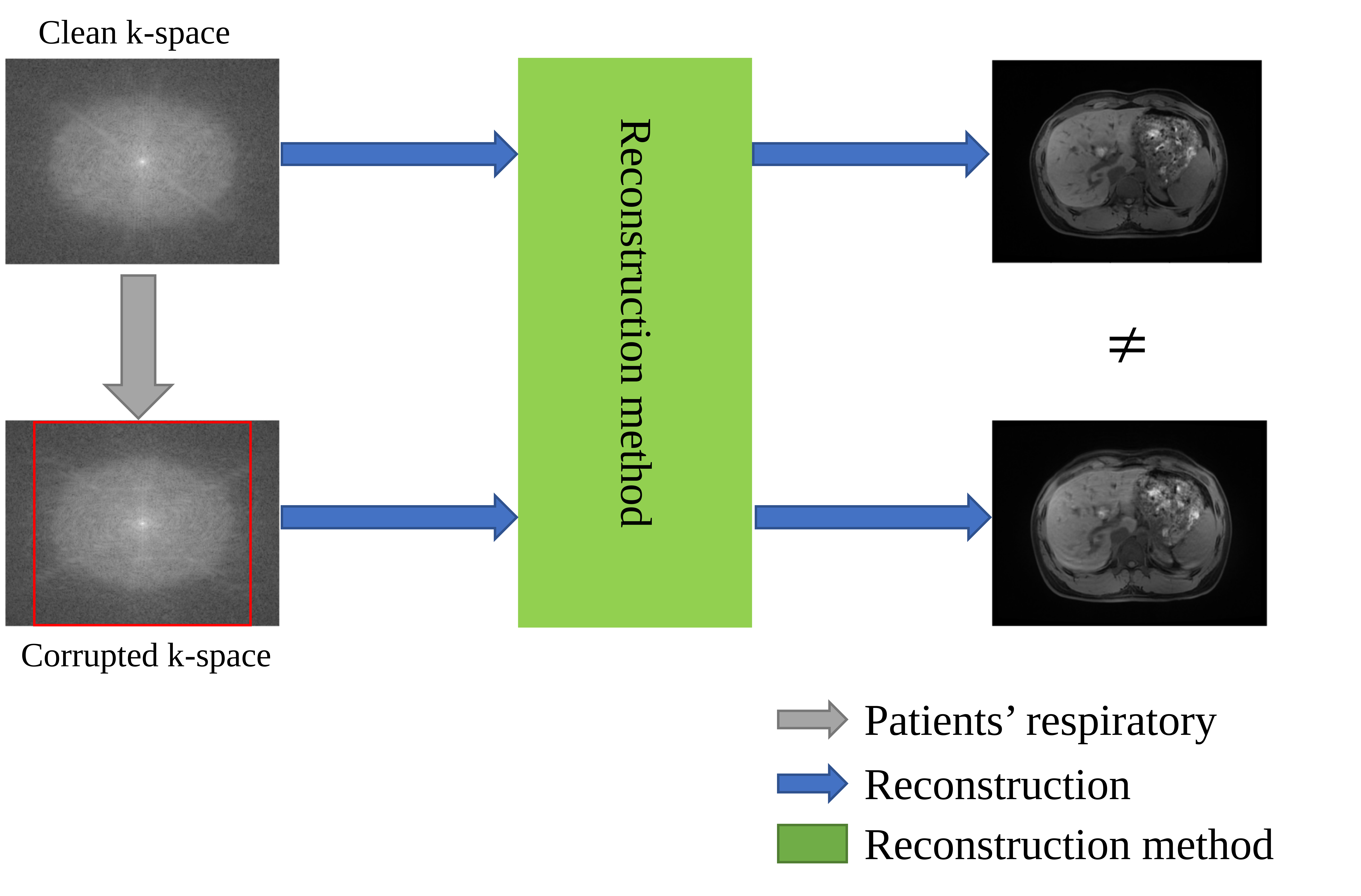}
    \caption{During the scanning period,patient's respiratory corrupted K-space data,while the relationship between K-space data before and after respiratory are nonlinear.It means that methods based on MR reconstruction can't remove artifact all. }
    \label{fig:recon}
\end{figure}

For respiratory movement, the orientation and position relative to FOV are not exactly the same as each phase encoding $k_y$. Using inverse Fourier transform of $s'(k_x,k_y)$ can only get images with artifacts from $\rho(x,y)$. As shown in Fig.\ref{fig:recon}, it means utilizing reconstruction-based method alone completely solve the problem of motion artifacts, requiring the incorporation of image-domain techniques.\par

\subsection{RAC as an I2IT task}
The mapping relationship from k-space to the image domain reveals that respiratory motion artifact is a global feature. Previous study\cite{tamada2020motion} modeled the RAC problem as an image restoration problem and used CNNs for solving it, but the performance in image restoration has been relatively weak. This is because CNNs primarily focus on local feature extraction and have limited capability in capturing global structural information. Although CNNs can increase their receptive field by stacking multiple convolutional layers, they still struggle to capture the overall global structure of the image. Some researchers have attempted to enhance the learning performance by fusing multi-level features through a multi-scale approach. However, in medical imaging, upsampling and downsampling may alter the physical properties of the image and introduce new artifacts.

Given the limitations of CNNs in learning global features, some researchers have proposed to address the RAC problem as an image-to-image translation (I2IT) task using generative models. Generative models are a class of machine learning models that aim to learn the underlying distribution of given data and generate new samples that exhibit similar characteristics and statistical properties to the original data. Generative models can leverage global contextual information and learn to perform image-to-image translation tasks. In contrast, traditional CNN-based methods primarily focus on local information. Generative models can utilize global contextual information to better understand and translate the overall image, thus preserving the image's structure and semantics during the translation process. 

\subsection{Reliable Generative Model}
 The mainstream approach to solving I2IT tasks is through the use of generative models. Generative models are a type of statistical machine learning models that aim to generate new samples that closely resemble the distribution and patterns observed in the training data. In other words, the goal of generative models is to approximate the parameter $p(\theta)$ of the training data distribution as closely as possible. Generative models can be categorized into explicit probability density models and implicit probability density models based on whether they explicitly compute the probability density function of the data. In previous research, implicit models, which are computationally efficient, have received more attention, with GANs being the most popular among them.

 The objective of GAN is to train a generator network $G$ to generate images that resemble the distribution of real data. During the training process, the generator and discriminator networks of GAN engage in a game, and the optimization objective can be formulated as follows:
\begin{equation*}
    \min_{G} \max_{D} V(D, G)
\end{equation*}
where $G$ is the generator and $D$ is the discriminator. When we perform gradient updates for both the generator and the discriminator simultaneously, we actually ignore the max operation when computing the gradient for the generator. In computing the gradient for the generator, we should treat $\max_{D} V(D, G)$ as the loss function for the generator and backpropagate through the maximization operation. However, backpropagating through the maximization operation is challenging and unstable, leading to the problem of mode collapse. It means generator produces only a subset of samples or repeatedly generates similar samples, failing to cover the entire distribution of training data\cite{bau2019seeing}. Moreover, in medical problems, there is a higher requirement for fidelity in generated images, especially in specific lesion areas and anatomical structures. However, GAN generates images by sampling from latent space and mapping them to the image space through the generator network, making it difficult to achieve precise control over the generated image features that may result in the loss of fine details and distortions in image features. Zhu ~\textit{et al.}\cite{zhu2017unpaired} introduces cycle consistency loss, which involves taking the transformed images and mapping them back to the original domain through the inverse transformation. However, it only aims to make the reconstructed images as close as possible to the original images.

In general, the process of generating samples in implicit models typically involves sampling from a latent space and mapping them through a generator network. However, due to the lack of an explicit probability distribution representation, implicit models often cannot provide precise control mechanisms to guide sample generation. This makes it difficult to achieve precise control over specific attributes or features of the generated samples during the generation process. In contrast, explicit models explicitly model the probability distribution of the data, allowing for precise control over the sample generation process. Additionally, by introducing conditional variables into the probability distribution, conditional sample generation can be easily achieved. This gives explicit models greater flexibility and applicability in tasks such as image translation. Consequently, more accurate and controllable sample generation can be achieved in the generation process. Therefore, for scenarios such as medical imaging that require higher fidelity, explicit generative models are a better choice than implicit generative models.

In explicit models, normalizing flow\cite{dinh2015nice,dinh2017density,kingma2018glow} is considered to have good expressiveness and stability. Combining variable transformations, normalizing flow transforms a simple prior distribution into a complex target distribution through a series of invertible transformations. Each transformation introduces a Jacobian matrix that measures the impact of the transformation on the probability density function. When applying an invertible transformation, the determinant of the Jacobian matrix represents the scale change of the volume in the data space before and after the transformation. By ensuring that the determinant of the Jacobian matrix is 1, the transformation does not introduce additional volume scaling or distortion, thus preserving the consistency of the data distribution.

\begin{figure}
    \centering
    \includegraphics[width=11.5cm]{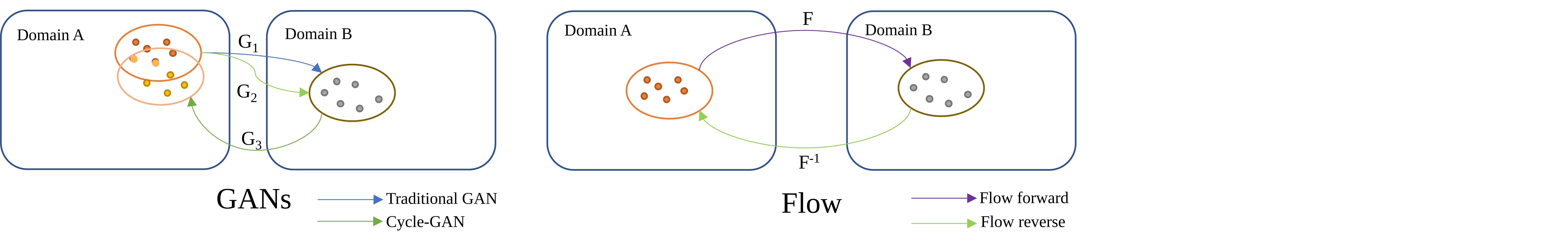}
    \caption{Compared to GAN, Flow is composed of a series of invertible mappings and excels in the accuracy of modeling distributions.}
    \label{fig:GANvsFlow}
\end{figure}

\section{Methodology}
In this section, we introduce an image restoration framework, named {\color{red}artifact-free flow restorer (AF$2$R)}, that learns the mapping from the distribution of artifact-corrupted image to clean image. We provide details of the framework in following.

\subsection{Nonlinear relationship between artifact and anatomy structure}
In previous studies\cite{tamada2020motion,lee2016acceleration}, some researchers defined the artifact removal problem as a traditional linear image restoration task, and the model is usually formulated as:
\begin{equation}
    I = J+R
\end{equation}
Where I and R represent anatomical structure and artifact respectively. Previous studies have treated respiratory artifact as addictive noise, and some progress has been made~\cite{tamada2020motion}. 
As illustrated above, it is inappropriate to consider artifact caused by respiratory motion as addictive noise. Instead, a nonlinear relationship needs to be introduced. Inspired by the screen blend technique in Photoshop, we define the relationship between the anatomical content and artifacts as:
\begin{equation}
    I = J+R-\lambda(J*R)
\end{equation}
Where $\lambda$ is the coupling coefficient which may be changed according to the number of iterations or depth of network and $*$ represents the matrix pointwise operation. Our defined model is more closely aligned with the process of encoding artifact information into the original signal compared to simple additive composite models. For instance, in regions with low signal intensity, the artifacts are more clear. When the respiratory motion is weak and there are fewer artifacts, the anatomical content is well presented.

\subsection{Arthitecture}
\begin{figure*}[ht]
\begin{center}
\includegraphics[width=18cm]{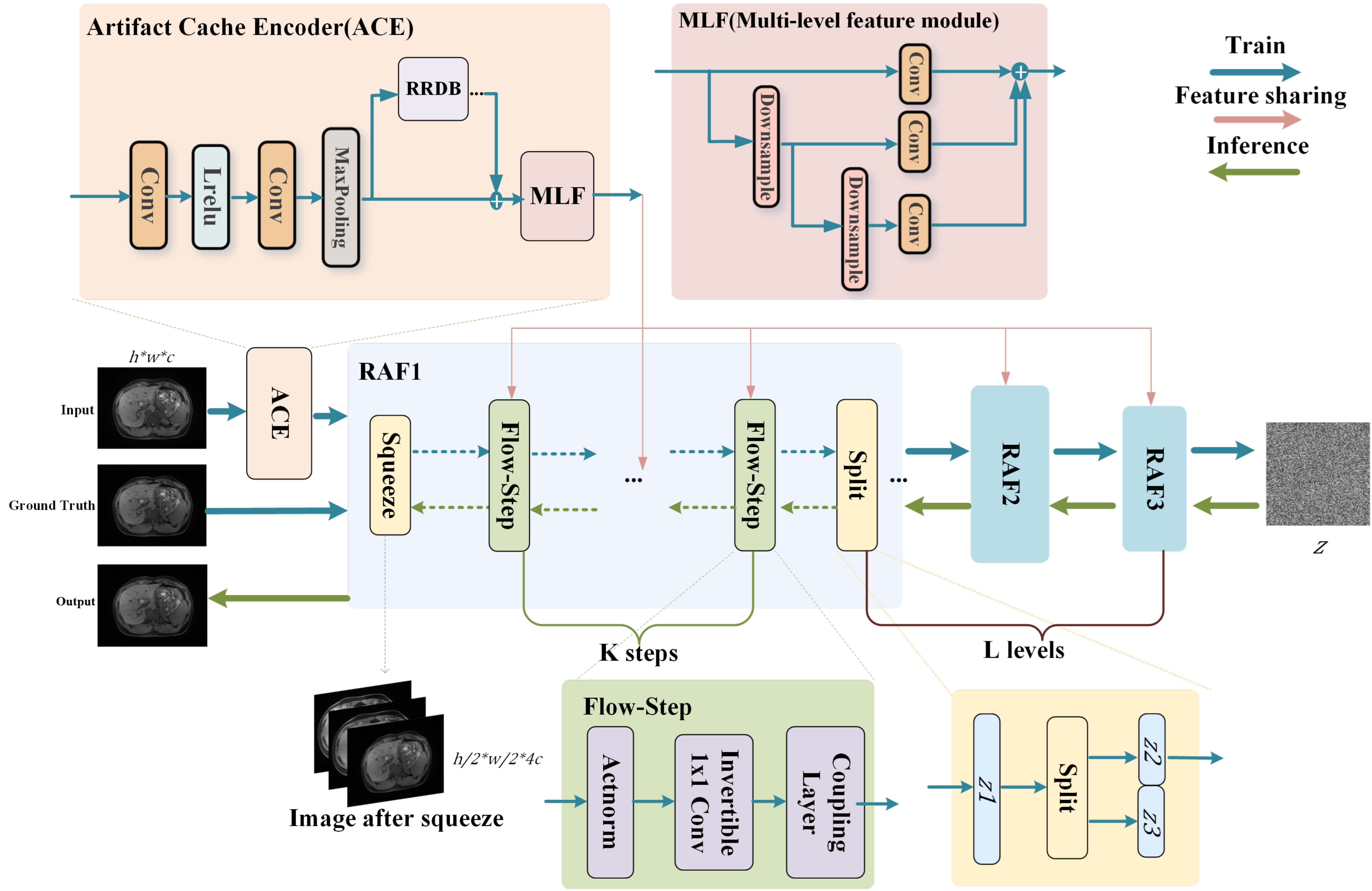}
\end{center}
\caption{Overall flow of the proposed method for motion artifact removal in image domain.}
\label{fig:Network_Artichture}
\end{figure*}

The overall architecture of AF$2$R is illustrated in Fig.~\ref{fig:Network_Artichture}. The proposed framework main consists of two components, including Artifact Cache Encoding (ACE) module $\mathbf{\mathcal{G}}$ and removal artifact flow-based (RAF) module $\mathbf{\mathcal{F}}$. Our network learns a mapping that transforms a artifact-corrupted image distribution $p(x)$ to the conditional artifact-free image distribution $p_{(y|x)}(y|x,\theta)$. 

\subsection{Artifact Cache Encoding module}
In order to obtain more effective image representations, we first encode the input images by an ACE module $\mathbf{\mathcal{G}}$, which is extract features from the artifact-corrupted images or artifact-free image. The popular Residual-in-Residual Dense Blocks (RRDB)~\cite{wang2018esrgan} is used for the backbone of ACE module $\mathbf{\mathcal{G}}$, which is to ensure training stability. Since our goal is to remove artifacts and there is no change in image resolution, we choose to remove the upsampling layers at the end of RRDB. To obtain richer image features for the pixel-level restoration problem, we also incorporate a multi-scale feature extractor  to obtain features at multiple levels (MLF)~\cite{dinh2017density}.  The detailed structure of ACE module $\mathbf{\mathcal{G}}$ is shown in Fig.~\ref{fig:Network_Artichture} (a) and (b).

\subsection{Removal Artifact Flow-based module}
Here, we introduce the content of the conditional flow module for artifact removal. Based on the analysis in the previous sections, we have confirmed that the relationship between artifacts and anatomical structures in the image domain is nonlinear. Traditional convolutional network structures cannot effectively capture this relationship. Inspired by this analysis, we introduce a new coupling layer in the following sections to enhance the flexibility and expressive power of the model, enabling it to better adapt to the artifact removal task.

Figure.\ref{fig:Network_Artichture} illustrates the proposed model architecture. The entire process can be described as the transformation from the conditioned features (extracted by the ACE module) to the latent space z. Similar to Glow, we adopt a multi-scale architecture with $L$ levels. Each level consists of three parts: squeezing operation, $K$ flow steps, and splitting operation. The task of flow steps is to gradually transform the input from the image domain to the latent space. Each flow step consists of three different layers: Actnorm layer, Invertible $1\times1$ convolution, and the coupling layer based on the artifact removal task.

Firstly, the input image undergoes a squeezing operation along the channel dimension to preserve the spatiotemporal correlation of the feature maps. It should be noted that since medical images are single-channel grayscale images, we need to make a  dimensionality expansion on the image before inputting it into the model. Through the squeezing process, we divide the input along the spatial dimensions into patches and directly reshape each block into $1\times1\times4c$, the size of the input feature map changes from $h\times w\times c$ to $\frac{h}{2}\times\frac{w}{2}\times4c$. Then, we normalize the feature map channel-wise using the actnorm layer instead of batch-normalization(bn)\cite{}, which holds the aim of alleviating activations noise caused by bn. Next, we use invertible $1\times1$ convolution to facilitate information interaction and mixing between different channels, which helps improve the expressive power and diversity of the features.

Many existing flow models increase the expressive power of the model and achieve stronger representation capabilities between the image space and the latent space by stacking multiple coupling layers. In previous works~\cite{dinh2015nice,dinh2017density}, Addictive coupling layers and affine coupling layers, belong to the category of linear transformations, are introduced to reduce the computational effort. However, the linear coupling layers are limited in their representational capability and also less suitable for downstream tasks of quality control. To capture the variation relationship between artifacts and anatomical details, we propose a new coupling layer based on the modeling results mentioned earlier:
\begin{equation*}
  h_{i} = m(x_{i-1})+x_{i}-\lambda(x_{i}*x_{i-1})
\end{equation*}
where * represents pointwise operation, and lambda represents the non-linear coefficient. We assume that the transformation between $x_{i}$ and $x_{i-1}$ simulates the actual artifact decomposition process, which achieves better artifact removal results with the stacking of multiple coupling layers.

Finally, we use the split operation to split the latent variables, providing the model with more control and generation options, thereby increasing the flexibility and expressive power of the model. Besides, through the Split operation, we divide the input along the channel dimension into two equally sized parts and send one of them to the next flow operation to save computation. In our experiments, $L$ is set to 3, and $K$ is set to 12.For AF$2$R, it has $L$ RAF module $\mathbf{\mathcal{F}}$, each RAF $\mathbf{\mathcal{F}}$ itself containing $K$ flow steps.

\subsection{Loss function}
Given a large set of training pairs $(X_i, Y_i)$ (where $i$ denotes the sample number) consisting of images with and without motion artifacts, we define the RAC problem as the task of learning the conditional probability distribution to transform artifact-corrupted images $X$ into artifact-free images $Y$. This is achieved by minimizing the negative log-likelihood (NLL) loss function:
\begin{eqnarray*}\label{nll loss}
\mathcal{L} (\theta;X_i,Y_i) \!\!\!\!\!&=&\!\!\!\!\!-\log p_{Y_i|g_{\theta}(X_i),\theta} \nonumber\\
                         &=&\!\!\!\!\!-\log p_z(f_{\theta}(Y_i;g_{\theta}(X_i)))-\log \left| \det \frac{\partial f_{\theta} }{\partial Y_i}(Y_i;g_{\theta}(X_i))\right|,
\end{eqnarray*}
where $g_{\theta}(\cdot)$ is ACE, and $\det$ represents the Jacobian determinant, illustrating the density transformation caused by the reversible network RAF.


\section{experiment}
In this section, we explore the performance of the proposed method on both simulated and real-world data containing artifacts, which show the effectiveness of our network architecture design and our network’s ability in removing motion artifacts. 

\subsection{Experimental Setup}
\subsubsection{Data Preparation}
We use three sequences for acquiring the real abdomen data from $16$ healthy volunteers on a $3$T scanner (United Imaging Healthcare, Shanghai, China), which include fat-suppressed sequence, dualEcho sequence and water-fat separable sequence. The corresponding parameters are setting as following. For fat-suppressed sequence and water-fat separable sequence, we using the following parameters: FOV $= 300 \times 400mm$, acquisition matrix $= 342 \times 456$, slice thickness $= 3mm$, {\color{red}TE/TR $= 3ms$}, acquisition time is set as $15s$, $26s$, and $46s$. The different setting of dualEcho sequence is acquisition matrix, which is set as $306 \times 408$.  In the experiment,  all the volunteers are able to hold their breath well, that we regarded these data as ground truth, named as data-$15s$. The data scanning with $25~s$ and $46~s$ acquisition time are in order to get data with real respiratory artifact, where volunteers can not hold breath, named as data-$25s$ and data-$46s$. All in vivo experiments were approved by the Institutional Review Board (IRB) of Shenzhen Institutes of Advanced Technology, and informed consent was obtained from each volunteer.

\subsubsection{Evaluation metrics}
To quantitatively evaluate the performance of {\color{red}AF$2$R} for RMAC, both Quantitative metrics and visual perception are needed. Several image quality metrics, including peak signal to noise ratio (PSNR)\cite{}, structural similarity index (SSIM)\cite{wang2004image}, {\color{red}visual information fidelity (VIF)\cite{} and universal quality index (UQI)\cite{}},  were utilized to gauge the quality of the artefact-corrected images, which can be measured as follows:
\begin{equation*}
    PSNR = 20*log_{10}\frac{max(GT)\sqrt{N}}{||GT-Result||},
\end{equation*}
and
\begin{equation*}
    SSIM = l(GT,result)*c(GT,result)*s(GT,result),
\end{equation*}
where $Result$ is the output image of models, $GT$ denotes the image which is artifact-free and $N$ is the total number of image pixels. We used the default settings for all the hyper-parameters of the evaluation metrics. For all metrics, higher values indicate better performance for removing artifact. And the anatomical details should be measured through visual perception.

\subsubsection{Compared methods} To verify the effectiveness and superiority of {\color{red}AF$2$R}, we compared it with some state-of-the-art works based on image domain that are closely related to our task, that is, latest GAN-based methods, including UnpairGAN~\cite{oh2021unpaired} and UGP~\cite{upadhyay2021uncertainty}, traditional image restoration methods based on CNN, MARC~\cite{tamada2020motion}. In addition, excellent work on the removal of CT metal artifact, ADN~\cite{liao2019adn}, is also included in the discussion.

\subsubsection{Implementation details}
In producing motion-corrupted images for train, we use the simulation method proposed in reference~\cite{shaw2020k} to generate the elastic motion artifacts. The data-$15s$ is choosed as the corresponding reference to construct paired data, while the parameters and effects that generated the artifact were evaluated by radiologists. All the images are center-croped to $256\times256$ for training. In all experiments, the Adam optimizer was used with parameters $(\beta_1, \beta_2) = (0.5, 0.999)$ and a learning rate of $10^{-4}$, and the patch size is set to $8$. Each experiments was trained until overfitting was observed or when reaching $100,000$ iterations. All experiments were conducted using PyTorch~\cite{paszke2019pytorch} on NVIDIA~Quadro~A6000~GPU~with~48GB~memory. 

\subsection{Ablation Study}
The goal of this section is to experimentally investigate different ablation studies that make the proposed network powerful enough to correct the motion artifacts from MR images. Every time, our flow-based models only change a hyper parameter or a function in the model. In particular, we majorly scrutinize on three vital element: the coupling layer, the coupling coefficient $\lambda$ and the depth of our model. All experiments were performed on synthetic data to facilitate qualitative and quantitative analysis of experimental effects.  

\subsubsection{Coupling layers}
\begin{figure}[ht]
\begin{center}
\includegraphics[width=9cm]{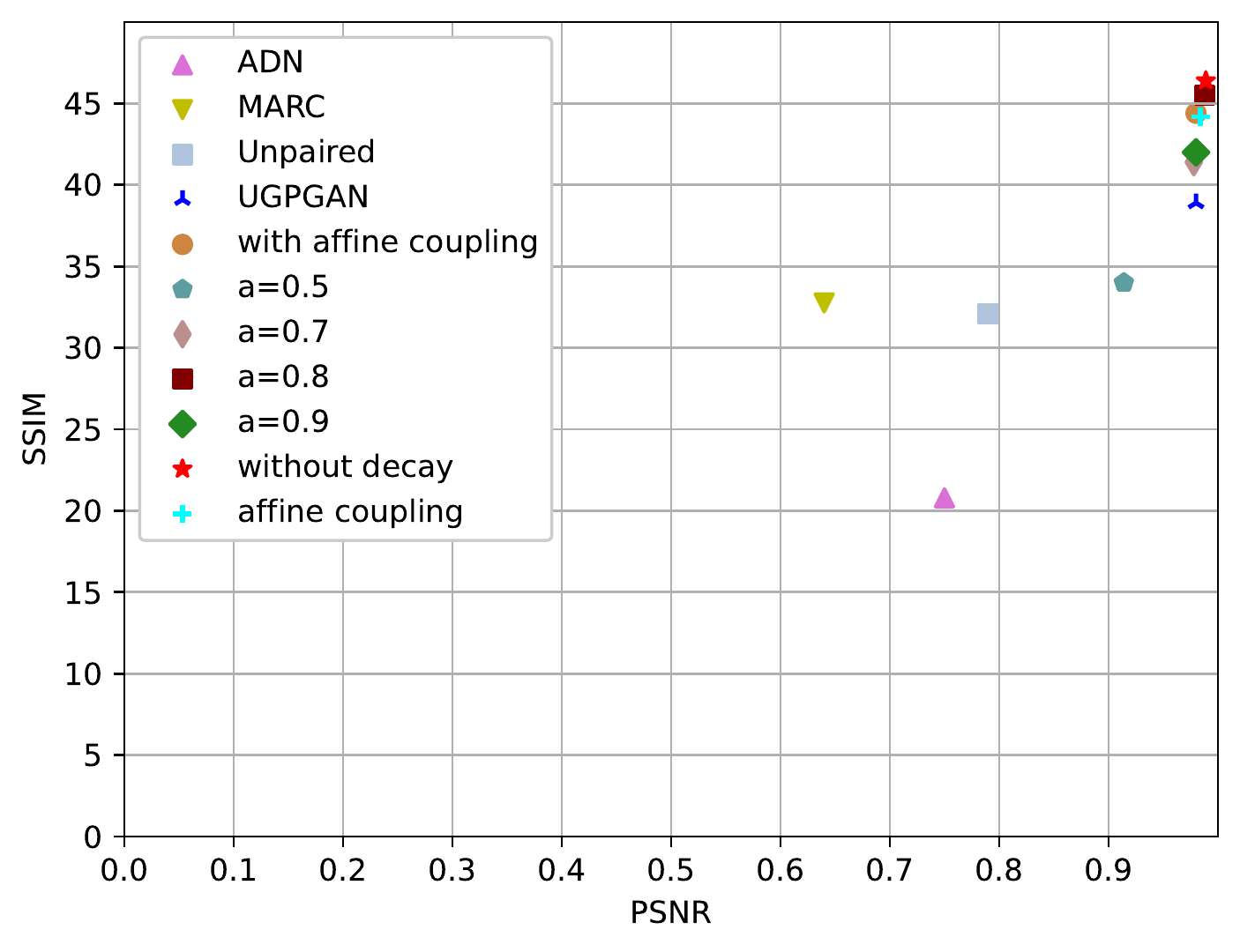}
\end{center}
\caption{The comparative result of the line and nonlinear coupling layer, which include the results of decay parameter.}
\label{fig:Coupling_layer_compare}
\end{figure}

To demonstrate the correctness of our approach in capturing the relationship between artifacts and anatomical content, we first provide an example. For comparison, we used a traditional flow network that utilizes affine coupling layers to illustrate the effectiveness of our designed coupling layer. And at the same time, we also analyse the decay of the coupling coefficient $\lambda$. Besides, aiming at exploring link of nonlinear strength between each flow step. So, we set a dacay model of $\lambda$ as $\lambda = a^k * \lambda$, where $a$ is the decay parameter of $\lambda$ and $k$ is the value of which layer of flow step.  The corresponding results is shown in Fig.~\ref{fig:Coupling_layer_compare}
 Under the same iteration rounds, the nonlinear coupling layer proposed is significantly better than the affine coupling layer in quantization index. This indicates that our proposed nonlinear relationship is closer to the relationship between artifact and anatomical structure than the traditional linear relationship.
 
\subsubsection{Flow step number}
With the intention of verifying the relation ship between the depth and performance of proposed model, we conducted experiments at different flow step number $K$. 
\begin{figure}[ht]
\begin{center}
\includegraphics[width=9cm]{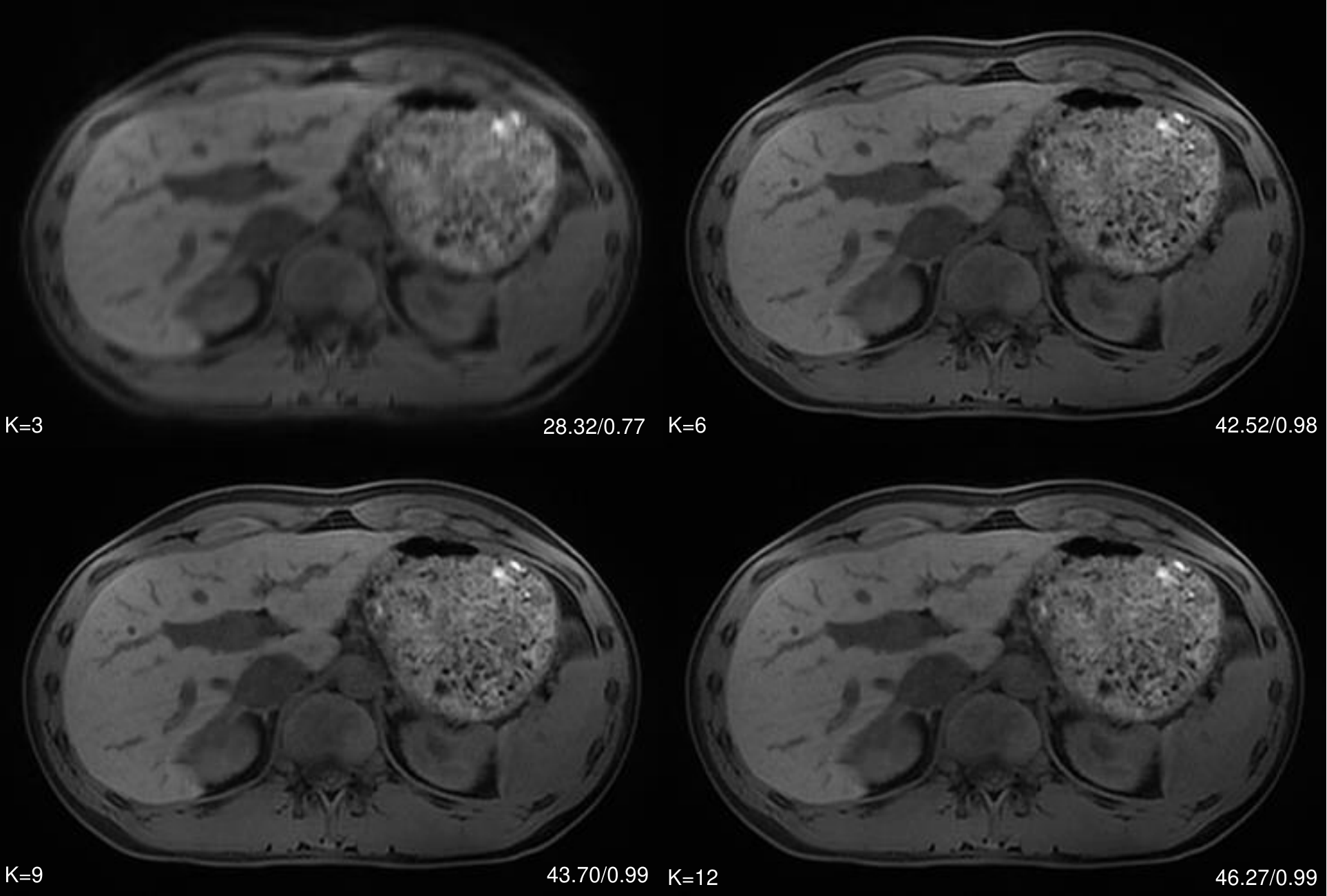}
\end{center}
\caption{The results of different flow step numbers.}
\label{fig:Flow_step}
\end{figure}

Fig.~\ref{fig:Flow_step} shows results on fs data. 
Flow is composed of multiple flow steps that enable powerful nonlinear fitting by coupling these steps. In this study, we compared the impact of varying the number of flow steps on the model's performance, as presented in Fig. \ref{fig:Flow_step}. Our findings revealed that, within a specific range of flow step numbers, increasing the number of flow steps led to improved quantitative metrics (such as PSNR and SSIM) and enhanced visual quality of the generated images. This observation further validates the effectiveness of our designed coupling layer in addressing the RAC problem.

\subsubsection{Model width}
To examine the influence of model width, we conducted training experiments with different numbers of hidden channels in two conditional layers, as depicted in Table 5. Reducing the number of hidden channels resulted in increased artifacts within complex structures, whereas increasing the number of hidden channels yielded improved motion artifact removal for MR. As a result, we established the number of hidden channels in our model as 64.
flow-steps K and hidden layers in two conditional layers.
\begin{table}[ht]
 \renewcommand{\arraystretch}{1.3}
 \caption{Result of Model width}
 \label{table_width}
 \centering
 \begin{tabular}{cccc}
  \hline
  \bfseries channel number & \bfseries PSNR & \bfseries SSIM\\
  \hline
   channel = 64 & 46.42 & 0.99\\
   channel = 32 & 28.74& 0.68\\
  \hline
 \end{tabular}
\end{table}

\subsection{Results on simulated data}
This section shows the motion artifacts correction performance of the proposed method in terms of quantitative index and visual quality on simulated data. 
\begin{figure*}[ht]
    \centering
    \includegraphics[width=18cm]{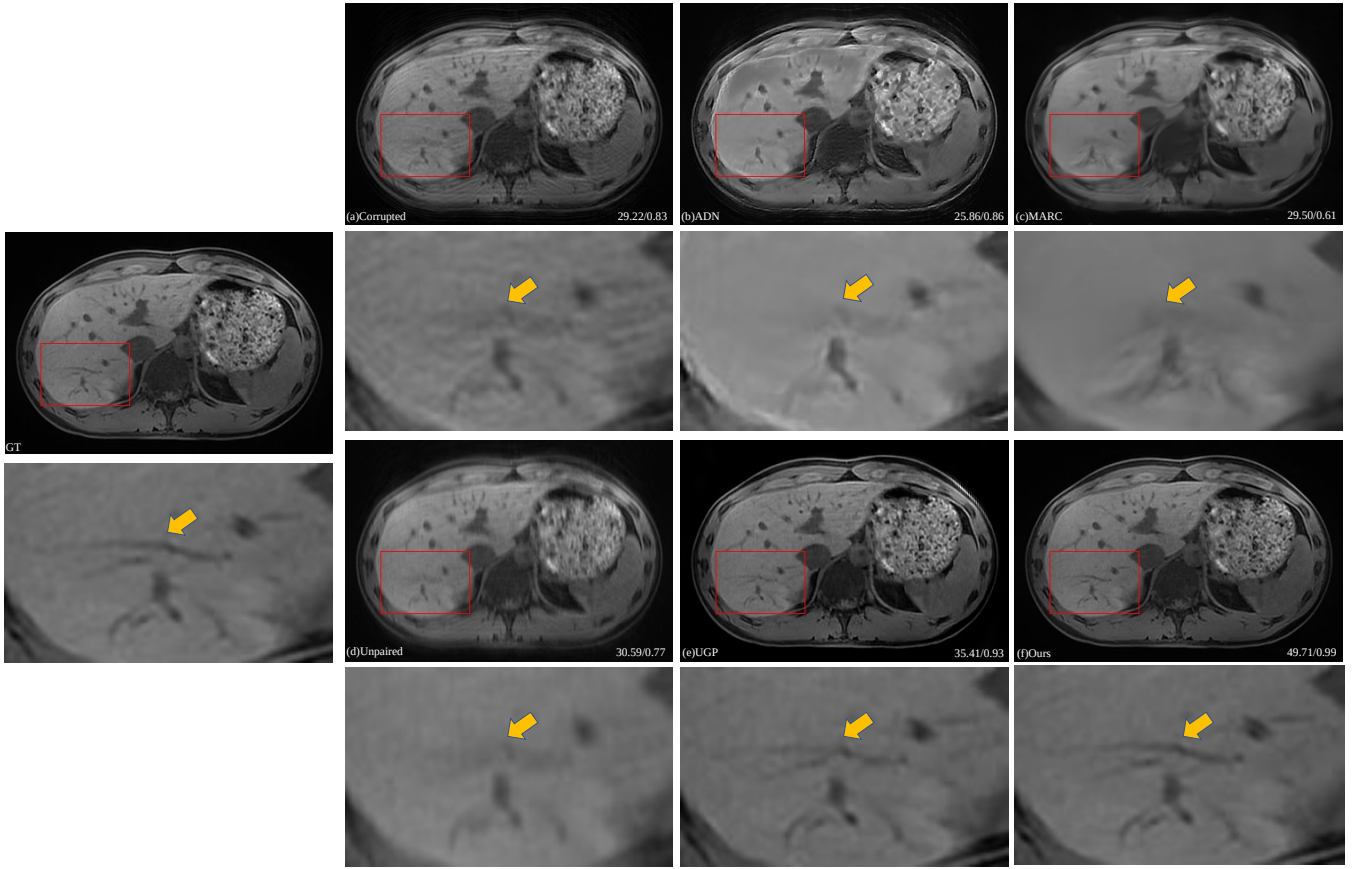}
    \caption{Respiratory artifact removal results of fat-suppressed sequence data using various parameters with simulated artifact: (a) Artifact corrupted, (b) ADN
(c) MARC, (d) Unpaired, (e) UGP, (f)Ours, (g)the ground truth}
    \label{fig:FS_simulated_results}
\end{figure*}
\begin{figure*}[ht]
    \centering
    \includegraphics[width=18cm]{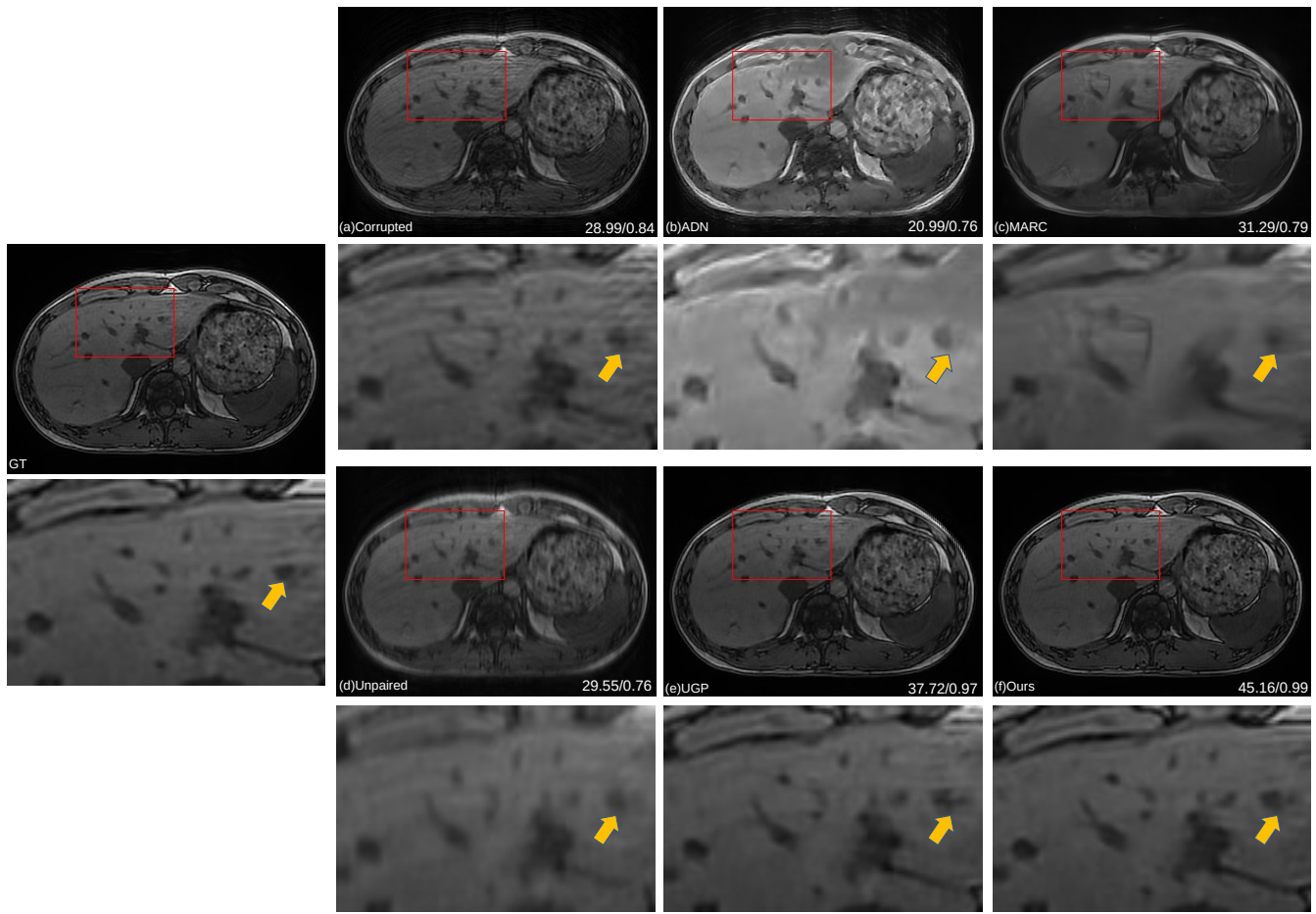}
    \caption{Respiratory artifact removal results of DualEcho sequence data using various parameters with simulated artifact:(a) Artifact corrupted, (b) ADN
(c) MARC, (d) Unpaired, (e) UGP, (f)Ours, (g)the ground truth}
    \label{fig:DE_simulated_results}
\end{figure*}

\begin{figure*}[ht]
    \centering
    \includegraphics[width=18cm]{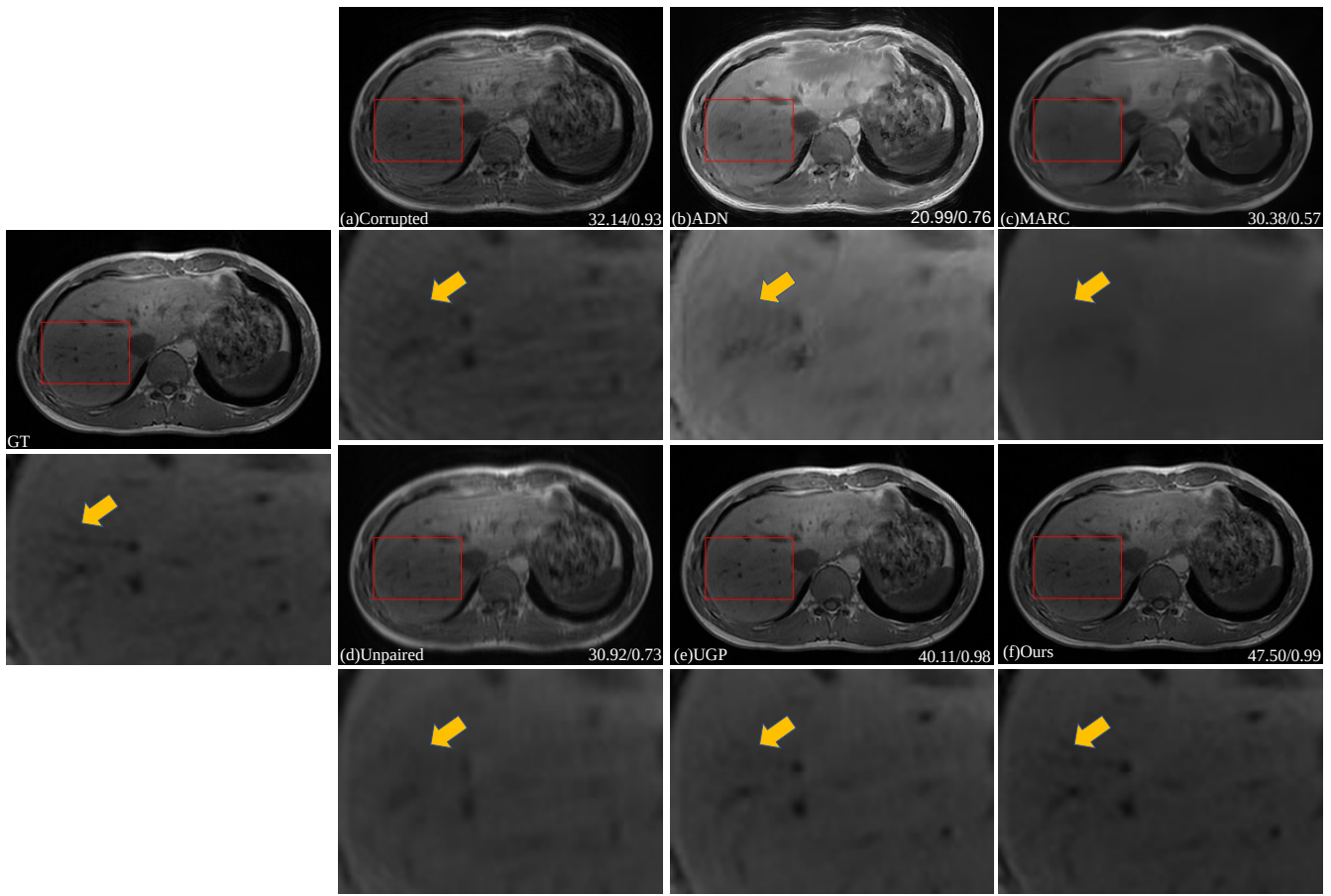}
    \caption{Respiratory artifact removal results of water-fat sequence data using various parameters with simulated artifact:(a) Artifact corrupted, (b) ADN
(c) MARC, (d) Unpaired, (e) UGP, (f)Ours, (g)the ground truth}
    \label{fig:WFI_simulated_results}
\end{figure*}



The quantitative comparison results of our proposed method and other methods on the simulation data are shown in Table \ref{table:Syn results}. It is evident that the generative model-based methods outperform the CNN-based methods in terms of PSNR and SSIM, indicating the advantage of data-driven modeling methods in the RAC problem. In particular, the uncertainty-guided progressive UGPGAN\cite{upadhyay2021uncertainty} performs better than other methods by incorporating uncertainty estimation and progressive training. With the introduction of the Flow model and improved coupling layers, our method further enhances the quantitative performance. Specifically, we achieve significant improvement in terms of PSNR compared to the second UGPGAN, and a slight improvement in terms of SSIM, demonstrating the effectiveness of our image-domain modeling approach in the RAC problem.

We further present the visual comparisons between our method and the comparative methods on the simulation data in Figure~\ref{fig:FS_simulated_results}, Figure~\ref{fig:DE_simulated_results} and Figure~\ref{fig:WFI_simulated_results}. The CNN-based method MARC\cite{tamada2020motion} removes the majority of artifacts but smoothens the overall image, thereby sacrificing some image details. In contrast, Unpaired\cite{oh2021unpaired} preserves more anatomical details but introduces image blurring, leading to a decrease in diagnostic quality. UGPGAN\cite{upadhyay2021uncertainty} achieves better visual results, but there are geometric distortions in certain anatomical details or the generation of content that does not exist in the original image, which is undesirable in clinical practice. In comparison, our method is capable of removing most streaking artifacts and effectively restoring the anatomical details, ensuring the highest possible fidelity of the reconstructed image.

\begin{table*}[ht]\label{table:Syn results}
\setlength\tabcolsep{3pt}
\begin{center}
\begin{tabular}{p{2.2cm}<{\centering}|*{2}{p{2.2cm}<{\centering}}|*{2}{p{2.2cm}<{\centering}}|*{2}{p{2.2cm}<{\centering}}} 
\toprule
\multicolumn{1}{c|}{\multirow{2}{*}{Methods}}&\multicolumn{2}{c|}{fat-suppressed sequence}&\multicolumn{2}{c|}{DualEcho sequence}&\multicolumn{2}{c}{water-fat separable sequence}\\ 
                                   & PSNR           & SSIM          & PSNR           & SSIM          & PSNR           & SSIM          \\ 
\midrule
Corrupted                          & 30.57          & 0.92          & 29.95          & 0.92          & 28.95          & 0.87          \\
ADN~\cite{liao2019adn}             & 21.92          & 0.83          & 21.51          & 0.85          & 20.43          & 0.77          \\
MARC~\cite{tamada2020motion}       & 31.72          & 0.83          & 31.66          & 0.81          & 30.42          & 0.78          \\
UnpairGAN~\cite{oh2021unpaired}    & 30.36          & 0.70          & 29.56          & 0.66          & 28.78          & 0.64          \\
UGP~\cite{upadhyay2021uncertainty} & 36.82          & 0.98          & 38.66          & 0.99          & 37.64          & 0.98          \\
Ours                               & \textbf{47.58} & \textbf{0.99} & \textbf{46.05} & \textbf{0.99} & \textbf{46.07} & \textbf{0.99} \\ 
\bottomrule
\end{tabular}
\caption{Quantitative comparison of various methods for simulated data.}
\end{center}
\end{table*}

\subsection{Result on real data}
Next, we evaluated the performance of our proposed method on real data, where there are no available ground truth, thus only
qualitative comparisons were performed. The qualitative evaluation
results on the real motion dataset are shown in Fig. \ref{fig:Real_results}. The red box represents the effect of artifact removal, while the green box represents the preservation of anatomical details. In the real scene, Unpair\cite{oh2021unpaired} shows better effects than other comparative experiments, with good artifact removal effect and better preservation of anatomical structures. However,the image quality after processing is relatively general,and the area with artifact is blurred after processing. For Marc\cite{tamada2020motion} and ADN\cite{liao2019adn}, it is still a problem in the simulated artifact scene,and the image details will be damaged when the artifact is removed. UGP\cite{upadhyay2021uncertainty} has a certain gap in the removal effect of real artifact compared with that of simulated artifacts, and it will introduce secondary artifact. Compared with simulated artifacts, real respiratory artifact have stronger randomness and weaker structure. The generalization of UGP is relatively general for this case.Compared with the above methods, our proposed method eliminates most obvious artifact and fringes, and does not introduce secondary artifact. Moreover, in terms of visual perception, our method well preserves image details without obvious image damage.
\begin{figure*}[htbp!]
    \centering
    \includegraphics[width=18cm]{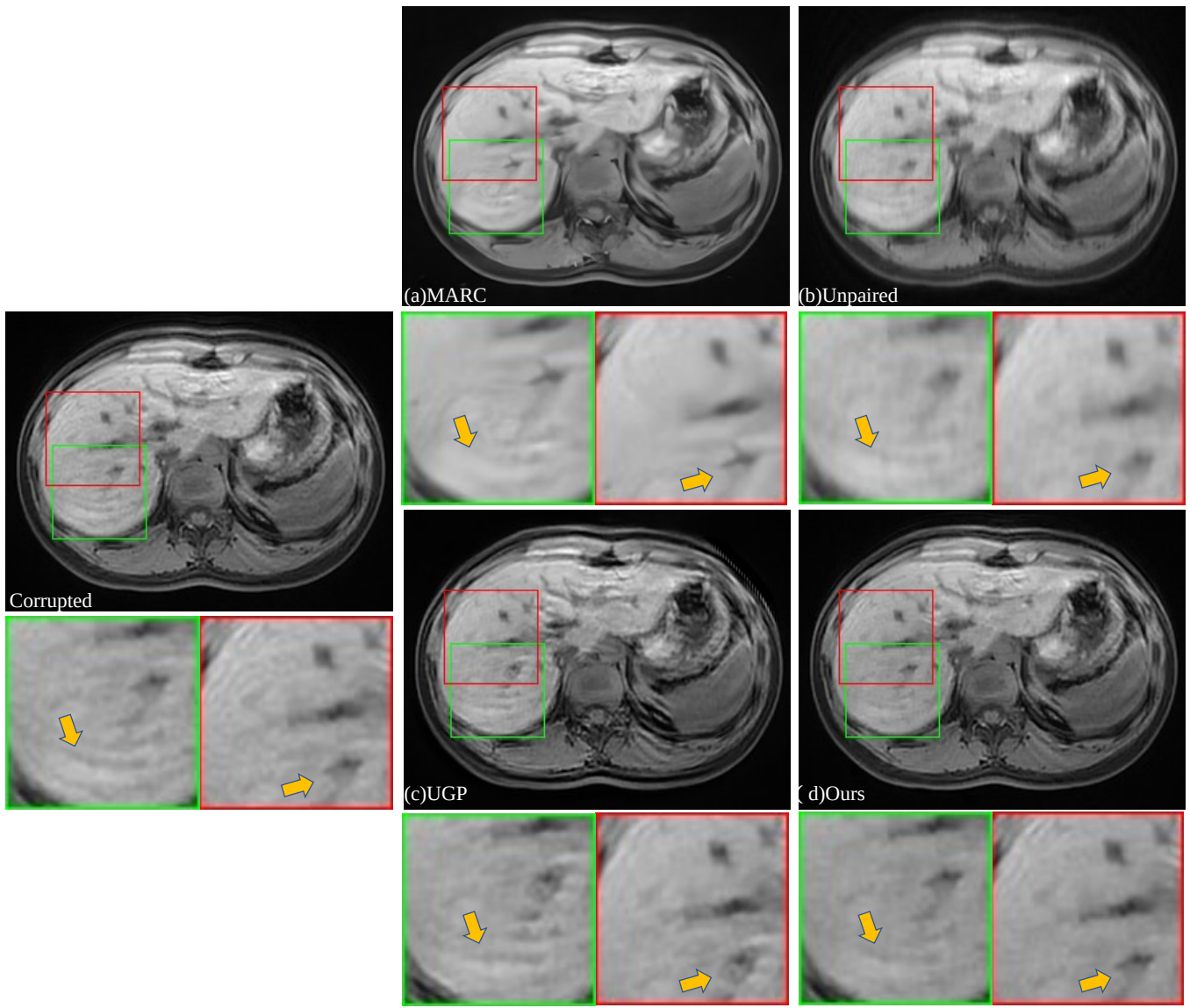}
    \caption{Respiratory artifact removal results of real respiratory artifact data:(a) Artifact corrupted, (b) ADN
(c) MARC, (d) Unpaired (e) UGP (f)Ours}
    \label{fig:Real_results}
\end{figure*}

\begin{figure*}[htbp!]
    \centering
    \includegraphics[width=18cm]{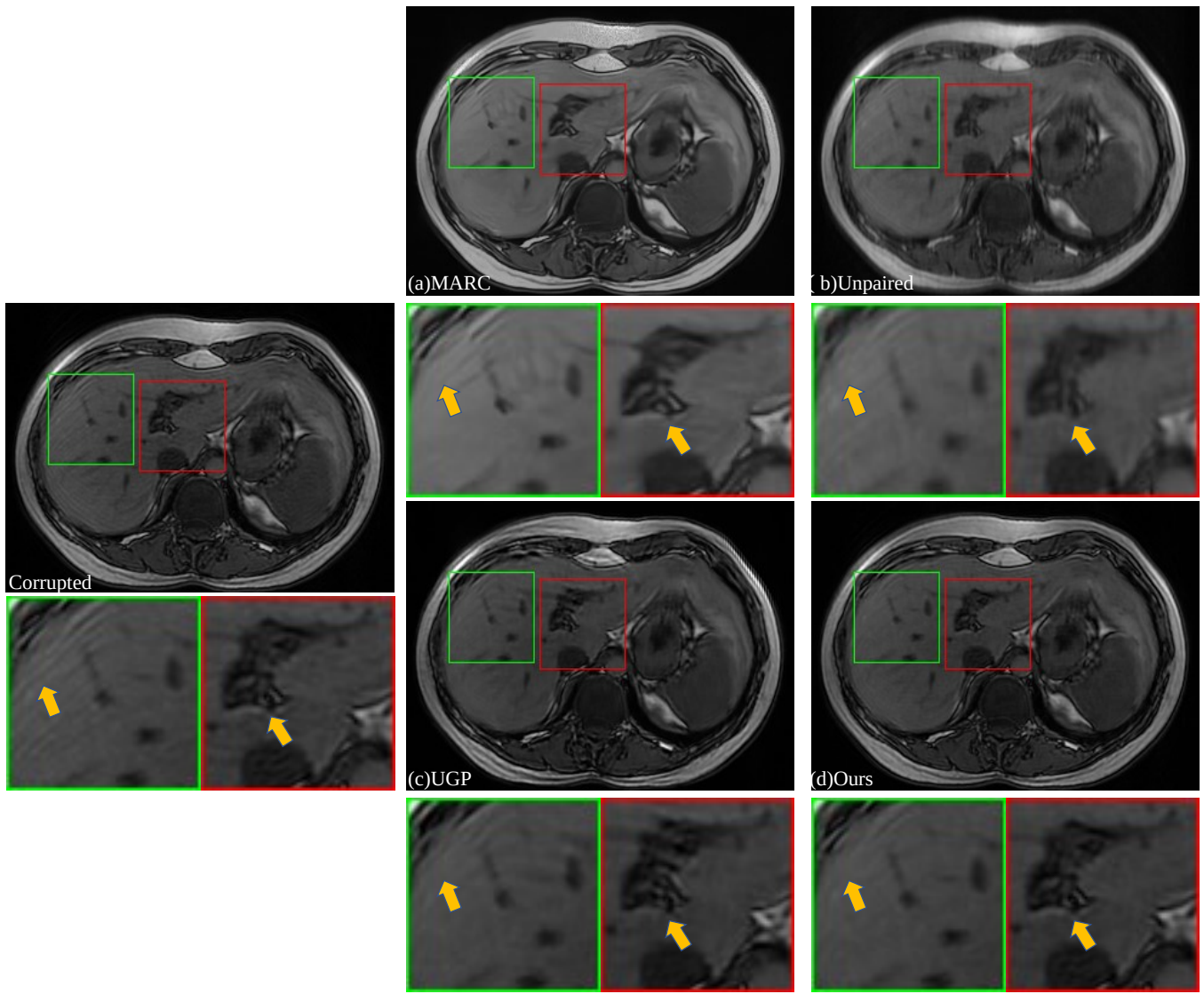}
    \caption{Respiratory artifact removal results of real respiratory artifact data:(a) Artifact corrupted, (b) ADN
(c) MARC, (d) Unpaired (e) UGP (f)Ours}
    \label{fig:Real_results}
\end{figure*}

\subsection{Performance Between implicit models and implicit models in medical scenarios}
From the perspective of artifact removal in simulated and real scenarios, GAN-based implicit generative models can partially remove artifacts but at the cost of losing some anatomical details or introducing distortions. Since it is not possible to directly compute the probability distribution of samples, accurate probability estimation of generated samples and evaluation of the model's quality becomes challenging. Therefore, the application of implicit models in medical imaging has certain limitations. In contrast, explicit generative models such as normalizing flows, which are based on explicit modeling, can directly model conditional distributions, allowing for more precise control over the generated sample's features while preserving anatomical structures better when removing artifacts.

In summary, implicit generative models are suitable for tasks where precise control over the generation process is not required, while explicit generative models are suitable for tasks that demand accurate control over generated samples and interpretability. Considering the medical scenarios we are facing, explicit generative models are undoubtedly the better choice.

\section{Conclusion}

In this paper, we explore the necessity of post-processing research in the image domain for motion artifact correction. Unlike previous studies, we redefine the relationship between the content of MR images and respiratory artifacts in the image domain. Taking a global perspective, we achieve artifact removal by constructing a conditional normalizing flow. Building upon this redefined model, we introduce a new nonlinear coupling layer. In contrast to most adversarial learning methods based on GANs, our approach learns the distribution of artifact-free images from images corrupted with artifacts, resulting in a more reliable method with improved image perception quality. Experimental results on synthetic data and real respiratory motion artifact data demonstrate that our method outperforms existing approaches in terms of quantitative metrics and visual perception quality. Beyond this task, we aim to generalize our method to other medical image restoration tasks or other medical imaging modalities.

\section*{Acknowledgment}
This work was supported in part by the National Natural Science Foundation of China under Grant 12001180, and Grant 12026603.

\bibliographystyle{IEEEtran} 
\bibliography{Refs}

\end{document}